\documentclass[11pt]{amsart}
\usepackage{amsbsy,amssymb,amscd,amsfonts,latexsym,amstext,delarray,
amsmath,graphicx}

\newtheorem{thm_def}{Theorem--Definition}[section]
\newtheorem{thm}{Theorem}[section]

\newtheorem{lem}[thm]{Lemma}

\newtheorem{rem}[thm]{Remark}

\numberwithin{equation}{section}

\def\C{{\mathbb C}}

\def\Z{{\mathbb Z}}
\def\R{{\mathbb R}}

\def\cD{{\mathcal D}}

\def\cF{{\mathcal F}}

\def\cH{{\mathcal H}}

\def\cL{{\mathcal L}}

\def\cQ{{\mathcal Q}}

\def\cU{{\mathcal U}}

\def\text{\hbox}

\title[Renormalization group in Epstein--Glaser scheme]{Deformations of quantum
field theories and Connes--Marcolli's renormalization group  in Epstein--Glaser scheme}

\author[Ceyhan]{\"Ozg\"ur Ceyhan}
\address{\"O.~Ceyhan: Korteweg-de Vries Institute for Mathematics, University of Amsterdam
P. O. Box 94248, 1090 GE Amsterdam, Netherlands} 
\email{o.ceyhan@uva.nl}

\begin{document}

\maketitle

\section{introduction}
Over the decades, physicists have developed a number of state of the art techniques to 
produce quantities of great physical relevance out of mathematically ill-defined
quantum field theories (QFTs).  The general strategy is to define physical quantities
in families in such a way that they  are well-defined  away from the certain limits of
the parameters, and then to extract finite limits of these quantities by using regularization
techniques.

The momentum space renormalizations have been widely used in QFTs. The 
combinatoric of divergencies and regularizations are beautifully encoded by  
Hopf algebras of Feynmann diagrams and Birkhoff decomposition of loops
in these Hopf algebras (see \cite{ck1,ck2}). The complexified dimension plays 
the role of deformation parameter in Connes--Kreimer picture. In   \cite{cm,cm1}, 
Connes \& Marcolli have constructed a Riemann--Hilbert correspondence 
associated to perturbative renormalization based on Connes--Kreimer's approach. 
They have observed an action of a pro-unipotent affine group scheme ${\mathbb U}^*$, 
universal with respect to the physical theories, and pointed out its connection to the 
motivic Galois group of the scheme of 4-cyclotomic integers $\Z[i][\frac{1}{2}]$.

On the other hand, Epstein--Glaser renormalization distinguishes itself among others: It produces finite 
QFTs from the very definition by choosing the domain of physical parameters suitably.  
The Epstein--Glaser's approach is based on  Dyson series 
\begin{eqnarray}
\label{eqn_s_matrix}
S = 1 + \sum_{n=1}^{\infty} \frac{(i)^n}{n!} \int_{M^{n}} dx_1 \cdots dx_n \ 
T_n(\cL_I(x_1),\dots,\cL_I(x_n)) 
\end{eqnarray}
of scattering operator (S-matrix) for a given potential term $\cL_I$ of a Lagrangian.
The problem in Epstein--Glaser setting is formulated as the problem of  extensions of 
distributions $T_n(\cL_I(x_1),\dots,\cL_I(x_n)) $ defined on the configuration
space $M^n \setminus \Delta$ of points on the spacetime $M$ to the diagonals
$\Delta$.

Contrary to the common perception, that points at divergencies as sources of ambiguities 
in QFTs,   ambiguities are still present in finite QFTs and are determined by distributions 
supported on the diagonals in Epstein--Glaser setting. This short note aims to describe
the deformations of QFTs in terms of the distributions supported on the diagonals and
then give an action of the pro-algebraic group ${\mathbb U}^*$ which appears in 
Connes--Marcolli's setting, on the finite QFTs constructed by Epstein-Glaser renormalization
scheme.

This paper is organized as follows: In Section 2, we review some basic facts on
Epstein--Glaser constructions of time ordered products. In the following section, we 
describe the deformation space of QFTs. In Section 4, we  give an action of the 
pro-algebraic group  ${\mathbb U}^*$ on the space of QFTs. Finally, in Section 5, we 
discuss a number of corollaries of our constructions in \S3 and \S 4, and their connections 
to some other renormalization related problems.

\section{Epstein-Glaser renormalization in a nutshell}
\label{sec_eg}
Let spacetime  $M$ be Euclidian space $\R^d$, and $\cD(M)$ be the space of test functions 
on $M$  with the usual topology. Let $\cH$ denote the Hilbert space of the free fields, 
$D$ a suitable dense subspace and $\Omega$ be the vacuum state.

\subsection{Time ordered products}
Time ordered products form a collection of operator valued distributions
\begin{eqnarray}
\label{eqn_t_product}
\{ T_N: \cD(M^{N}) \to End(D) \mid N:= \{1,\dots, n\}\},
\end{eqnarray}
and, in Epstein-Glaser renormalization scheme, they are expected to satisfy a 
set of basic properties:

\subsubsection{Symmetry} $T_N$'s are symmetric under permutations of indices, i.e.,
$$
T_N(f_1 \otimes \dots \otimes f_n) = T_N(f_{\sigma(1)} \otimes \dots \otimes f_{\sigma(n)})
$$
for all $\sigma$ in the symmetric group of index set $N$.

\subsubsection{Causality} $T_N$ factorizes casually, i.e.,  if $I, I^c \ne \emptyset$ is 
a partition of  $N$, and if $supp (f_i) \cap supp (f_j) = \emptyset$ for all $i \in I$ and $j \in I^c$,
then
\begin{eqnarray}
\label{eqn_causal}
T_N(f_1 \otimes \dots \otimes f_n) = T_I (\bigotimes_{i \in I} f_i)  \cdot  T_{N \setminus I}  (\bigotimes_{j \in I^c} f_j). 
\end{eqnarray}

\subsubsection{Translation invariance}$T_N$ is invariant under translations:
$$
T_N(f_1(x_1) \otimes \dots \otimes f_n(x_n)) =  T_N(f_1(x_1-a) \otimes \dots \otimes  f_n(x_n-a)) . 
$$

Epstein \& Glaser  constructed time ordered products essentially by using the causality in \cite{EG}.

\begin{thm}[Epstein \& Glaser, \cite{EG}]
Time-ordered products exist.
\end{thm}

\subsection{Wick expansions of time ordered products}
The extension problem of operator valued distributions is reduced to an extension problem for 
numerical distributions by expanding time ordered products in terms of the Wick expansions. 

\begin{thm}
Let $:\phi^{k_1}(x_1):, \dots, :\phi^{k_n}(x_n):$  Wick mononomials
for noncoinciding points $x_1,\dots,x_n$ in  $M$. Then
\begin{eqnarray}
T_N(:\phi^{k_1}(x_1): \cdots :\phi^{k_n}(x_n):)  =  \sum_{\mathbf{J}=(i_1,\dots,i_n) = 0}^{(k_1,\dots,k_n)}
t_{\mathbf{J}}(x_1,\dots,x_n)
\times \frac{:\phi^{i_1} \cdots \phi^{i_n}:}{i_1! \cdots i_n!} \nonumber
\end{eqnarray}
where the numerical distribution $t_{\mathbf{J}}(x_1,\dots,x_n)$ is 
$$
\langle \Omega, T_N(:\phi^{k_1-i_1}(x_1): \cdots :\phi^{k_n-i_n}(x_n):) \ \Omega \rangle
$$
\end{thm}
(for instance, see Theorem 2.4 in \cite{BurFred}).

\section{Deformations of QFTs in Epstein--Glaser scheme}
One of the main consequences of Epstein--Glaser construction   is that the
space  which  parameterizes  the collection of time ordered products can be observed 
explicitly:

\begin{lem}
\label{lem_r_group}
Let $T_N, \widehat{T}_N$ be two different time ordered products. Then, the difference 
$T_N - \widehat{T}_N$  is an operator valued distribution supported on the diagonals 
$\Delta \subset M^{N}$.
\end{lem}

The proof of this lemma is straightforward  and can be found in \cite{pin} and \cite{kel}  
for Minkowski and Euclidean cases respectively.

\subsection{The space of QFTs}
\label{sec_space}
We can reformulate  Lemma \ref{lem_r_group} on the level of numerical distributions 
and  give the space of QFTs as follows: First, we use the  translation invariance and 
set one of the points, say $x_1$, to $0$, so that $t_{\mathbf{J}}\in \cD(M^{|N|-1})$ for 
$n \geq 2$. Due to Lemma  \ref{lem_r_group}, we obtain a new distribution  by adding 
another numerical distribution supported on the union of diagonals 
$\Delta = \bigcup_{I \subset N}  \Delta_I$  where 
$\Delta_I := \{ (0,x_2,\dots,x_n) \mid  x_i=x_i \ \text{iff} \ i,j \in I \} \subset M^{|N|-1}$, i.e., 
\begin{equation}
\label{eqn_shift}
t_{\mathbf{J}} \mapsto t_{\mathbf{J}} + d_{\mathbf{J}}, \ \ \  \text{where} \ \ \  
d_{\mathbf{J}}= \sum_{I \subset N} d_{\mathbf{J},I}, 
\end{equation}
and $d_{\mathbf{J},I}$'s are  numerical distributions supported on  the corresponding diagonals
$\Delta_I  \subset M^{|N|-1}$.

Due to the well known fact that distributions supported at one point are finite linear 
combinations of the $\delta$ distribution and its derivatives, the summand supported
on the deepest diagonal $\Delta_N = \{0\}$ in (\ref{eqn_shift}) is given by
$$
d_{\mathbf{J},N}=\sum_{{ \alpha= (\alpha_1,\dots, \alpha_{nd})}  
\atop { \sum \alpha_* \leq sd(t_{\mathbf{J}})}} 
b_{\mathbf{J},N}^\alpha \cdot 
\frac{\partial^{\alpha_2} }{\partial x_2 ^{\alpha_2}}  \cdots  
\frac{\partial^{\alpha_{nd}} }{\partial x_{nd}^{\alpha_{nd}}}  \delta_{N}
$$
where $\delta_{N}$ is the delta function supported on $\{0\} \subset M^{|N|-1}$.
The degree is  bounded by the  generalized degree of homogeneity, called scaling
degree 
$$
sd(t_{\mathbf{J}}) :=  \inf \{s \mid \lim_{\lambda \to 0} \lambda^s  \cdot \int\ t_{\mathbf{J}}(\lambda x) 
\omega(x) dx \}.
$$
The other $ d_{\mathbf{J},I}$'s in (\ref{eqn_shift}) can be given as above case;
\begin{equation}
\label{eqn_expansion}
 d_{\mathbf{J},I}= \sum_{{ \alpha= (\alpha_1,\dots, \alpha_{nd})}  
\atop { \sum \alpha_* \leq sd(t_{\mathbf{J}})}} 
b_{\mathbf{J},I}^\alpha \cdot  
\frac{\partial^{\alpha_2} }{\partial x_2 ^{\alpha_2}}  \cdots  
\frac{\partial^{\alpha_{nd}} }{\partial x_{nd}^{\alpha_{nd}}}  \delta_{I}
\end{equation}
where $\delta_{I}$ is the delta function supported on $\Delta_I \subset M^{|N|-1}$.

Hence, we can rephrase Lemma \ref{lem_r_group} as a deformation theory for QFTs
in Epstein--Glaser setting: Let  $\text{Def} (\cQ)$ be the space of QFTs  around a given 
QFT determined by the set of  numerical distributions $\cQ=\{t_\mathbf{J}  \}$. 

\begin{thm_def}
$\text{Def} (\cQ)$  is an infinite dimensional Euclidean space whose
coordinate  ring $\cH$ is $\C[b_{\mathbf{J},I}^\alpha]$ where  $|\mathbf{J}|>2, 
|\alpha| \leq sd((t_{\mathbf{J}}))$ and $I \subset N$. 
\end{thm_def}

It is important to note that $\text{Def} (\cQ)$ is unobstructed since the coordinate  ring 
is $\C[b_{\mathbf{J},I}^\alpha]$, and therefore all $k$-th order deformations 
extends to the next  order for all $k$.

\subsection{Filtration of $\text{Def} (\cQ)$}
$\text{Def} (\cQ)$ is filtered
$$
\emptyset \subset \text{Def}^{(1)} (\cQ) \subset \dots \subset  \text{Def}^{(n)} (\cQ)
\subset   \text{Def}^{(n+1)} (\cQ) \subset \cdots  
$$
according to the cardinality of index set $\mathbf{J}=(j_1,\dots,j_n)$:
$$
\text{Def}^{(n)} (\cQ) = \{d_{\mathbf{J}} \mid  |\mathbf{J}|  = n+1 \}.
$$
for all $n=1,2,3,\dots$ The inclusions $\iota: I \hookrightarrow N$ of index sets induce
imbeddings $\iota_\#: \text{Def}^{(|I|-1)} (\cQ) \hookrightarrow \text{Def}^{(|N|-1)} (\cQ)$.

\section{Renormalization group in Epstein--Glaser scheme}

\subsection{Symmetries acting on the space of QFTs}
The most general form of symmetries of $\text{Def} (\cQ)$ are given by the  pseudo-group 
of all formal (local) diffeomorphisms $\mu: \text{Def} (\cQ) \to \text{Def} (\cQ)$. More elaborate
symmetries form a Lie pseudo-group. They are prescribed by systems of  nonlinear 
equations on jet bundles that are satisfying  formal integrability and local 
solvability conditions. The remarkable fact is that the Maurer--Cartan form produces an 
explicit  form of the pseudo-group structure equations, see \cite{olv}. 

A version of such a symmetry group, called the group of diffeographisms, is introduced 
as  the group of formal diffeomorphisms tangent to the identity of the space of coupling 
constants of the theory  by Connes \& Kreimer in \cite{ck2}.

\subsection{Connes--Marcolli's renormalization group in Epstein--Glaser setting} 
\label{sec_rep}
From physics perspective, the subgroup of symmetries of $\text{Def} (\cQ)$ generated by 
the scaling transformations is of particular interest since it essentially gives the 
renormalization group. 

In this paragraph, we present the action of a subgroup of scalings on the space of QFTs. 
Namely, we consider a pro-algebraic group of the form  ${\mathbb U}^*  = {\mathbb U} \rtimes 
{\mathbb G}_m$  whose unipotent part  ${\mathbb U}$ is generated by scaling transformations
and is associated to the free graded Lie algebra $\cF(1,2,\cdots)_\bullet$ with one generator 
$e_n$  in each degree $n>0$. The semi-direct product is given by the grading of 
${\mathbb U}$.

\subsubsection{Pro-unipotent part ${\mathbb U}$} 
Consider  the  scaling transformations
\begin{equation}
b_{\mathbf{J},I}^\alpha    
\mapsto  
 \sum_{K \subseteq I} 
  \epsilon(\alpha,K,\mathbf{J}) \ b_{\mathbf{J},K}^\alpha   
\label{eqn_scaling}
\end{equation}
where
\begin{equation}
\label{eqn_matrix}
 \epsilon(\alpha,K,\mathbf{J}) = \left\{
\begin{array}{ccc}
\lambda^{|K|}  & \text{if}  & j_1 > j_2  \\
1  & \text{if}  & j_1=j_2 \\
0  & \text{if}  &  j_1 < j_2.
\end{array}
\right.
\end{equation}

They act upon the degree $n$ piece  $\text{Def}^{(n)} (\cQ)$ of $\text{Def} (\cQ)$. Note that, 
the definition of $\epsilon$ guarantees that the matrix in (\ref{eqn_scaling}) is upper triangular 
and therefore the transformation in (\ref{eqn_scaling}) is  pro-unipotent. 

The infinitesimal  generator of   (\ref{eqn_scaling}) is given by the following vector field
\begin{equation}
\mathbf{e}_{n} = 
\sum_{ \mathbf{J}: j_1 > j_2} \sum_{I \subset N, \alpha} 
\left(
 \sum_{K \subseteq I}   |K| b_{\mathbf{J},K}^\alpha
\frac{\partial}{\partial 
b_{\mathbf{J},K}^\alpha}
\right)  
\nonumber
\end{equation}
in $T^* \text{Def}(\cQ)$.
The pro-unipotent part ${\mathbb U}$ is associated to free graded 
Lie algebra $\cF(1,2,\cdots)_\bullet$ which is generated by the elements $e_n$ at each 
positive  degree $n$.  

\subsubsection{Multiplicative group $ {\mathbb G}_m$ and semi-direct product}
Consider the 1-parameter group of automorphisms
$$
\theta_z: d_{\mathbf{J}} \mapsto e^{nz}  \cdot d_{\mathbf{J}}, \ \  \ \forall z \in \C 
$$
implementing the grading. Its infinitesimal generator is given by the grading operator
$$
Y(d_{\mathbf{J}}) :=  \frac{d}{dz} (\theta_z d_{\mathbf{J}}) \mid_{z=0} = n\cdot d_{\mathbf{J}}. 
$$

Finally, we define, for all $u \in {\mathbb G}_m$, an action $u^Y$ on 
${\mathbb U}$ by
$$
u^Y(X) = u^n X, \ \forall \  X \ \text{of degree} \ n. 
$$
We can then form the semi-direct product
$$
{\mathbb U}^* = {\mathbb U} \rtimes {\mathbb G}_m
$$
and this shows that 

\begin{thm} \label{thm_main}
The pro-algebraic group  ${\mathbb U}^*$ acts upon the space $\text{Def} (\cQ)$. 
\end{thm}

${\mathbb U}^*$ is  universal with respect to the set of physical theories,
in the sense that it is canonically defined and independent of the physical theory.

\begin{rem}
In their seminal paper \cite{cm}, Connes \&  Marcolli considered the same ${\mathbb U}^*$
as renormalization group. Their motivation is to identify the renormalization group as a 
motivic Galois group as Cartier suggested in \cite{car}.  In their approach, the pro-unipotent 
part is the graded dual of the universal enveloping  algebra
$$
\cU(\cF(1,2,\cdots)_\bullet)^\vee
$$
as a Hopf algebra. 
Then, they showed that the Tannakian category of flat equisingular connections 
which they have obtained from the differential system of counterterms is equivalent to a 
category of representations of the affine groups scheme ${\mathbb U}^*$. 
\end{rem}


\section{Remarks and further directions}

There are numerous connections between Epstein--Glaser  and other approaches to
QFTs.  Below,  we summarize a few direct corollaries of the discussions of the previous 
section and speculate on a few possible applications in related fields.

\subsubsection{Spacetime other than Euclidean spaces}
In \S \ref{sec_rep}, we have presented an action of pro-algebraic group ${\mathbb U}^*$ 
on the space $\text{Def} (\cQ)$ of $S$-matrices on Euclidean spacetime. However, the 
basics of this approach can be directly adapted for any spacetime manifold $M$. Once 
the time  ordered products are given in terms of numerical distributions (as in \cite{BurFred}, 
for instance), one can define a representation of ${\mathbb U}^*$ by considering scaling 
properties of distributions supported on the diagonals $\Delta \subset M^n$. A
construction of time ordered products for  curved space-time along with  a discussion of 
renormalization group which is very close to our desciption here can be found in \cite{holw}.

\subsubsection{Causal treatment of gauge theories}
In their papers \cite{kre} and \cite{wal1,wal2}, Kreimer and van Suijlekom extended the 
results of Connes \&  Marcolli to gauge field theories by  discussing  the Slavnov-Taylor 
identities for the couplings at the Hopf algebra level. Van Suijlekom showed that  the 
Slavnov-Taylor identities  generate a Hopf ideal of Hopf algebra of Feynman diagrams. 
Hence, these identities are compatible with renormalization, and the affine group scheme 
${\mathbb U}^*$ remains  as a part of the renormalization picture. 

On the other hand, using Epstein--Glaser in gauge field theories is not new to physics 
literature and has been studied extensively in both abelian and non-abelian gauge theories 
(for instance, see  \cite{dhks,hol, hur,gri,sch}).  
The gauge invariance condition in the causal approach is expressed in every order of 
perturbation theory separately by a relation of the $n$-point distributions $T_N$ with the 
charge $Q$, the generator of the free operator gauge transformations
$$
[Q,T_N]= d_Q T_N, 
$$
and this equation essentially encodes Slavnov-Taylor identities. By using the gauge
invariant distributions supported on the  diagonals, one gives the role of symmetry group
of perturbative gauge theories to same ${\mathbb U}^*$. The essential 
tool for casual approach in gauge theories is time ordered products in Grassmann variables
and it can be found in \cite{sch}. A very detailed exposition for Yang-Mills theoy
can be found in \cite{hol}.

\subsubsection{Epstein-Glaser vs. dimensional regularization}
Even though, our main theorem states an action of the same pro-algebraic
group ${\mathbb U}^*$ on perturbative 
QFT's as in Connes--Marcolli's case, the action is quite different in nature. The main distinction 
is that ${\mathbb U}^*$ acts upon the counterterms in their case. By contrast, the representation in 
\S \ref{sec_rep} is given by an action on the renormalized values. 

Moreover, in Connes--Marcolli's construction, the affine group scheme ${\mathbb U}^*$ appears
as a motivic Galois group. However, it is unclear to us whether  ${\mathbb U}^*$ has any direct 
motivic  role in Epstein--Glaser renormalization in the form discussed above.  This question simply 
arises from the fact that the integrals in (\ref{eqn_s_matrix}) contain distributions not rational functions, 
and therefore they are not periods.

\subsubsection{Feynmann motives and motives of configuration spaces} \label{sec_rem_mot}
There is an ongoing search for motivic origins of Feynman amplitudes (for an extensive account, see 
\cite{mar} and reference therein). This is a program initiated by Kontsevich's suggestion that these
numbers should be related to mixed Tate motives. There are several positive and negative results
in this direction. 

Epstein--Glaser approach hints a motivic treatment of Feynman integrals: Feynman rules
associates a distribution to each Feynman graph on a configuration space which is also determined
by the same graph. The divergencies of these integrals can be treated by the techniques that
we have used for time ordered products. Alternatively,  one can use Fulton--MacPherson type of 
compactifications of these configuration spaces and try to obtain regularized integrals on them. 
Fulton--MacPherson  compactifications of these configuration spaces are mixed Tate motives when 
the spacetime  is itself mixed Tate. This observation is quite intriguing and we are planning to discuss 
this approach in a subsequent paper.

\subsubsection*{Acknowledgements}
I take this opportunity to express my deep gratitude to K.~Kremnizer and M.~Marcolli.
This paper essentially contains what I have learned from them. I also wish to thank to 
S.~Agarwala and W.~van Suijlekom for their interest and suggestions which have been 
invaluable for me.

Part of this work was carried out during the author's stay at Caltech, which he thanks for 
the hospitality and support. The author is partially supported by a NWO grant.



\begin{thebibliography}{99}


\bibitem{BergKr} C.~Bergbauer, D.~Kreimer, {\it The Hopf algebra of rooted trees in Epstein-Glaser 
renormalization}. Ann. Henri Poincar\'e 6 (2005), no. 2, 343Ð367.

\bibitem{BurFred} R.~Brunetti, K.~Fredenhagen, {\it  Microlocal analysis and interacting quantum 
field theories: renormalization on physical backgrounds}. Comm. Math. Phys. 208 (2000), no. 3, 623--661.

\bibitem{car} P. Cartier, {\it A mad dayÕs work: from Grothendieck to Connes and Kontsevich. 
The evolution of concepts of space and symmetry.} Bull. Amer. Math. Soc. (N.S.) 38 (2001), no. 4, 389Ð408.

\bibitem{ck1} A.~Connes, D.~Kreimer, {\it Renormalization in quantum field theory and the Riemann--Hilbert 
problem. I. The Hopf algebra structure of graphs and the main theorem.} Comm. Math. Phys. 210 (2000), 
no. 1, 249--273.

\bibitem{ck2} A.~Connes, D.~Kreimer, {\it Renormalization in quantum field theory and the 
Riemann--Hilbert problem. II. The $\beta$-function, diffeomorphisms and the renormalization group.} 
Comm. Math. Phys. 216 (2001), no. 1, 215--241.

\bibitem{cm} A.~Connes, M.~Marcolli,  {\it From physics to number theory
via noncommutative geometry II: Renormalization, the Riemann-Hilbert
correspondence, and motivic Galois theory}. in `Frontiers in Number Theory,
Physics, and Geometry, II' pp.617--713, Springer Verlag, 2006.

\bibitem{cm1} A.~Connes, M.~Marcolli,   {\it Noncommutative Geometry, Quantum Fields 
and Motives},Colloquium Publications, Vol.55, American Mathematical Society, 2008.

\bibitem{dhks} M.~D\"utsch, T.~Hurth, F.~Krahe, G.~Scharf, {\it Causal construction of Yang-Mills theories. I.} 
Nuovo Cimento A (11) 106 (1993), no. 8, 1029--1041.

\bibitem{EG} H.~Epstein, V.~Glaser, {\it The role of locality in perturbation theory.} 
Ann. Inst. Henri Poincar\'e, Section A, Vol. XIX, n. 3 (1973) 211.

\bibitem{gri} D.R.~Grigore, {\it  The standard model and its generalizations in the Epstein--Glaser approach to 
renormalization theory.} J. Phys. A 33 (2000), no. 47, 8443--8476. 

\bibitem{hol}  S.~Hollands, {\it Renormalized quantum Yang-Mills fields in curved spacetime.} 
Rev. Math. Phys. 20 (2008), no. 9, 1033--1172.

\bibitem{holw} S.~Hollands, R.M.~Wald, {\it  On the renormalization group in curved spacetime.}
Comm. Math. Phys. 237 (2003), no. 1-2, 123--160.

\bibitem{hur} T.~Hurth, {\it  Nonabelian gauge theories: the causal approach.}  Ann. Physics 244 (1995), no. 2, 
340--425.

\bibitem{kel} K.J.~Keller {\it Euclidean Epstein-Glaser Renormalization}, arXiv:0902.4789

\bibitem{kre} D.~Kreimer,  {\it Anatomy of a gauge theory.}
Ann. Physics 321 (2006), no. 12, 2757--2781. 

\bibitem{mar} M.~Marcolli, {Feynman motives}, World Scientific, 2010.

\bibitem{olv} P.J.~Olver,  J.~Pohjanpelto,  {\it Maurer-Cartan forms and the structure of Lie pseudo-groups.} 
Selecta Math. 11 (2005) 99-126.

\bibitem{pin} G.~Pinter, {\it Finite renormalizations in the Epstein Glaser framework and renormalization 
of the $S$-matrix of $\Phi^4$-theory}. Ann. Phys. (8) 10 (2001), no. 4, 333--363.

\bibitem{sch} G.~Scharf,  {\it Finite Quantum Electrodynamics: The Causal Approach},
Springer; 2nd edition, 1995. 

\bibitem{wal1} W.D.~Van Suijlekom, {\it The Hopf algebra of Feynman graphs in quantum electrodynamics.}
 Lett. Math. Phys. 77 (2006), no. 3, 265--281.

\bibitem{wal2} W.D.~Van Suijlekom,  {\it Renormalization of gauge fields: a Hopf algebra approach.}
 Comm. Math. Phys. 276 (2007), no. 3, 773--798. 



\end{thebibliography}
\end{document}